\documentclass[useAMS,usenatbib]{mn2e}
\usepackage{epsfig}
\usepackage{color}

\newcommand{\Msolar}{\mbox{\,$\rm M_{\odot}$}}        
\newcommand{\Lsolar}{\mbox{\,$\rm L_{\odot}$}}        
\def\simge{\mathrel{\raise1.16pt\hbox{$>$}\kern-7.0pt
  \lower3.06pt\hbox{{$\scriptstyle \sim$}}}} 			 
\def\simle{\mathrel{\raise1.16pt\hbox{$<$}\kern-7.0pt
  \lower3.06pt\hbox{{$\scriptstyle \sim$}}}} 			 

 \newcommand{\iso}[2]{\mbox{$^{#1}{\rm #2}$}}           

\title[]{White-dwarf red-giant mergers, early-type R stars, J stars and lithium}
\author[]{Xianfei Zhang$^{1}$\thanks{E-mail:
xiz@arm.ac.uk} and C. Simon Jeffery$^{1,2}$\thanks{E-mail: csj@arm.ac.uk}\\
$^1$Armagh Observatory, College Hill, Armagh BT61 9DG, UK\\
$^2$School of Physics, Trinity College Dublin, Dublin 2, Ireland
}

\begin{document}

\date{Accepted . Received ; in original form }

\pagerange{\pageref{firstpage}--\pageref{lastpage}} \pubyear{2011}

\maketitle

\label{firstpage}

\begin{abstract}
Early-type R stars and J stars are a special type of carbon star, having
enhanced nitrogen ($\rm [N/Fe]\approx 0.5$), lithium, a low
\iso{12}{C}/\iso{13}{C} ratio ($<15$) and no s-element enhancements. The merger
of a helium white dwarf  with a red giant is regarded to be a
possible model for the origin of early-type R stars, but the details of
nucleosynthesis are not clear. In this paper we investigate three possible
channels for helium white-dwarf + red-giant mergers, and find that, amongst the three, only a
high-mass helium white dwarf subducted into a low core-mass red giant can make an early-type
R star. Nucleosynthesis of elements carbon, nitrogen, oxygen and lithium
correspond well with the observations. Furthermore, we find that the J stars may
represent a short and luminous stage in the evolution of an early-R star.

\end{abstract}


\section{Introduction}
Carbon stars are characterized by $\rm C/O > 1$ in their atmospheres. They are
generally classified as spectral types N, R and J. The N-type carbon stars are
recognized as normal carbon stars in which carbon is enhanced by third dredge-up
during the asymptotic giant-branch (AGB) phase. R-type carbon stars are further
classified into two groups \citep{Shane1928}: early-R (hot) and late-R (cool).
Late-R stars are similar to normal carbon stars (N-type), and the s-element over
abundance is possibly produced during the AGB phase \citep{Zamora2009}. Early-R
stars are different from late-R stars; they have near solar metallicity and show
enhanced nitrogen ($\rm [N/Fe]\sim 0.5$), a low \iso{12}{C}/\iso{13}{C} ratio
($<15$) and no s-element enhancement \citep{Dominy1984,Zamora2009}. Also, the
early-R stars are lithium-rich and none are found in binary systems. The J-type
carbon stars are similar to early-R stars, but show a high luminosity, a smaller
ratio of \iso{12}{C}/\iso{13}{C} and more lithium on the surface
\citep{Abia2000}. The surface composition of N-type and late-R carbon stars can
be reproduced by low-mass AGB nucleosynthesis models \citep{Zamora2009}. But the
origins of the early-R stars and J stars are not clear, especially the
nucleosynthesis responsible for their surface chemistries. The fact that,
except for the wide binary BD$+02^{\circ}4338$, all early-R stars are
single stars, encouraged \citet{Izzard2007} to argue for an origin based on
binary-star mergers, which was supported by binary-star population-synthesis
calculations

According to \citet{Izzard2007},
a possible evolution channel for early-R stars is for a helium white dwarf (He WD) to
merge with a red giant-branch (RGB) star. As the helium white dwarf comes into
contact with the expanding red giant, a common envelope forms. Assuming
spiral-in occurs faster than the envelope can be ejected, the white dwarf will
merge with the helium core of the giant. The final outcome will be a single star
having a core comprising the sum of the helium white dwarf and the red-giant (RG)
helium core surrounded by an H-rich mantle. \citet{Izzard2007} discuss the
composition of the envelope, which seems to be able to reproduce an enriched
carbon surface. However, \citet{Piersanti2010} point out that the enhancement of
carbon depends on whether efficient helium burning occurs during the merger
process. To analyze the details of the merger process, \citet{Piersanti2010}
performed a three-dimensional smoothed-particle hydrodynamics (3D SPH)
simulation and discussed the subsequent evolution. According to their calculation, no
efficient helium burning occurred which would dredge up carbon-enriched material
to the surface. However, some high-mass helium-white-dwarf merger channels,
{\it i.e.} where the white dwarf mass is greater than 0.2\,\Msolar, were
not studied by \citet{Piersanti2010}.

In order to survey a wider range of models, we have simplified the SPH
simulation into a 1D stellar evolution calculation. A detailed nucleosynthesis
analysis is included. Our result shows that the merger of a high-mass helium
white dwarf with a red giant could be a possible evolution channel for early-R
stars and J stars.



\section{Methods}

To investigate evolution following the merger of two white dwarfs, \citet{Zhang12a}
considered three sets of assumptions concerning the rate of accretion onto the more
massive star, representing different distributions of fast and slow accretion. To
simulate the merger of a helium white dwarf with a red giant, following the results of
SPH simulations \citep{Yoon07,Loren09,Piersanti2010}, we consider only the case of fast
accretion (or a fast merger).

Numerical simulations of stellar evolution are carried out using the stellar-evolution
code Modules for Experiments in Stellar Astrophysics (MESA version 4028) \citep{paxton11}.
For our experiments, it is difficult to control the required final mass of
the helium white dwarf from a full binary star evolution calculation. Thus, an artificial
method is adopted. We start with a zero-age main-sequence star of mass 2.0$\Msolar$
(metallicity $Z=0.02$) and evolve it until the helium core reaches the required mass.
Then the hydrogen envelope is completely removed to produce a naked helium core,
essentially a pre-WD model. This is achieved artificially in MESA by switching off the
nucleosynthesis and applying a high mass-loss rate until the required model is obtained.
These models cannot ignite central helium burning and evolve
straight to the white-dwarf (WD) track, ending up with a luminosity $\log L/\Lsolar =-2$
(see \citet{Zhang12a} for details). Similarly, a model for the RG core is constructed in exactly
the same way, except that we consider separately the cases where the RG core is either
warm or cold. Here, a "warm" core represent a true model RG core which is only just degenerate
(degeneracy parameter $\eta \approx 3 (0.2 \Msolar) - 10 (0.3 \Msolar)$),
whilst a cold core is much more degenerate ($\eta = 12 (0.2\Msolar) - 80 (0.4 \Msolar) $).
Our models do not include rotation; all mixing is complete and driven by convection
(no semi-convection, no thermohaline mixing, no diffusion).
We separate the subsequent simulation into three steps.\\
\paragraph*{Step one:} merge the helium white dwarf with the helium core of the red giant.
Following \citet{Yoon07,Loren09} and \citet{Piersanti2010}, a fast accretion rate of
$10^4 \rm \Msolar\ yr^{-1}$ is adopted, implying that the merger takes a few minutes.  \\
\paragraph*{Step two:} Switch off time-dependent
terms, including nucleosynthesis. Numerically, build up a hydrogen envelope on top of the merged
helium core from step one. This is achieved by  reversing the process used to remove the envelope, in this case by
applying a negative mass-loss rate. The composition of the hydrogen envelope is defined to be identical to that
of the RG progenitor. The structure of the post-merger model is thus a cool degenerate
central helium core with a helium-rich hot shell surrounded by a hydrogen envelope.\\
\paragraph*{Step three:} Switch on time-dependent terms and evolve the post-merger
model. Stop evolution before the thermally-pulsing AGB phase starts.
No stellar winds were included in any of the
calculations. The post-merger models are {\it defined} to have a final mass of 2.0\Msolar; since the
initial red giant has a mass of 2.0\Msolar\ to which is added the WD mass,
we assume that a fraction of the
total has been ejected during the common-envelope phase, but have chosen this fraction arbitrarily
for the time being. One calculation with a final mass of 1.5\Msolar\ was made for comparison (\S
4.3.2).

\section{Models}
\begin{figure}
 \centering
 \includegraphics [angle=0,scale=0.6]{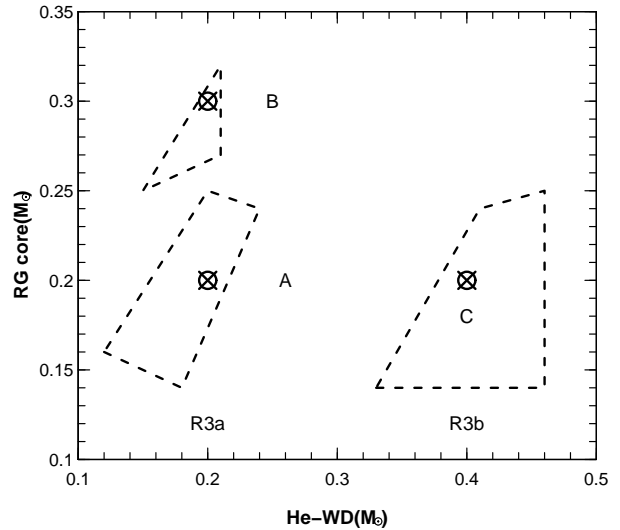}
 \caption{The circles with crosses shows the three types of model computed. Dashed lines indicate three zones defining possible evolution channels described by \citet{Izzard2007}.}
 \label{model}
 \end{figure}

According to \citet{Izzard2007}, there are two main WD-RG merger channels: low-mass He WD + RG (R3a) and high-mass He WD + RG (R3b). Channel R3a can be further divided into two groups: high core-mass RG and low core-mass RG. Thus, we set up three models for these three channels labelled Models A,  B and C respectively (see Fig.~\ref{model} for details).

\begin{figure}
 \centering
 \includegraphics [angle=0,scale=0.6]{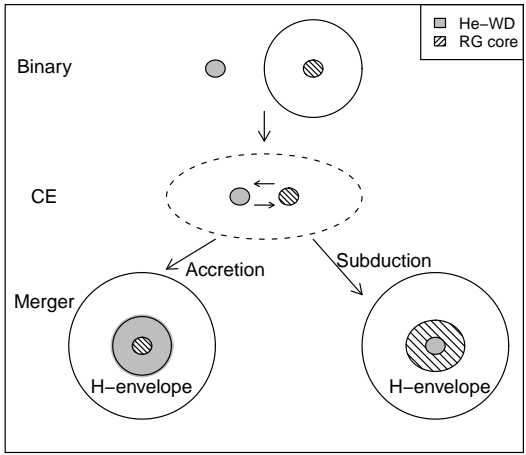}
 \caption{Cartoon illustrating two possible outcomes of a white-dwarf plus red-giant merger.}
 \label{pic}
 \end{figure}

In the \citet{Piersanti2010} SPH simulations, all the models are low-mass He WD
+ low core-mass RG mergers, and all the accretion is by mass transfer from the
helium white dwarf to the RG core. However, one concern is that the He WD has a
higher density than the RG core. The massive He WD may therefore subduct (or
sink) into the low-mass helium core. To cover these possibilities, we have
considered two separate merger types; \romannumeral1) accretion from the He WD
onto the helium core and \romannumeral2) accretion from the helium core onto the
He WD, i.e. the He WD subducts into the center of helium core; see
Table~\ref{table:models} and Fig.~\ref{pic} for details. As noted above, the
composition of the hydrogen envelope is taken from a 2\Msolar\ RG model which
has a 0.2\Msolar\ helium core, and the final masses are 2\Msolar.

\begin{table*}
\caption{Summary of the numerical experiments described in the text, showing the mass of the helium white dwarf,
the core mass of the initial red giant, the assumed final mass of the merger product, and the type of white-dwarf-core interaction. } 
\begin{tabular}{l*{3}{c}r}
\hline
 Models              & WD mass (\Msolar)& Core mass (\Msolar)& Final mass (\Msolar)& Merger type \\
 \hline
 Model A                   & 0.2 & 0.2 & 2.0 & accretion/subduction\\
 Model B1                  & 0.2 & 0.3 & 2.0 & subduction\\
 Model B2                  & 0.2 & 0.3 & 2.0 & accretion\\
 Model C1                  & 0.4 & 0.2 & 2.0 & accretion\\
 Model C2                  & 0.4 & 0.2 & 2.0 & subduction\\
 \hline
 \end{tabular}
 \label{table:models} 
\end{table*}

\section{Results and Discussion}

\begin{figure}
 \centering
 \includegraphics [angle=0,scale=0.6]{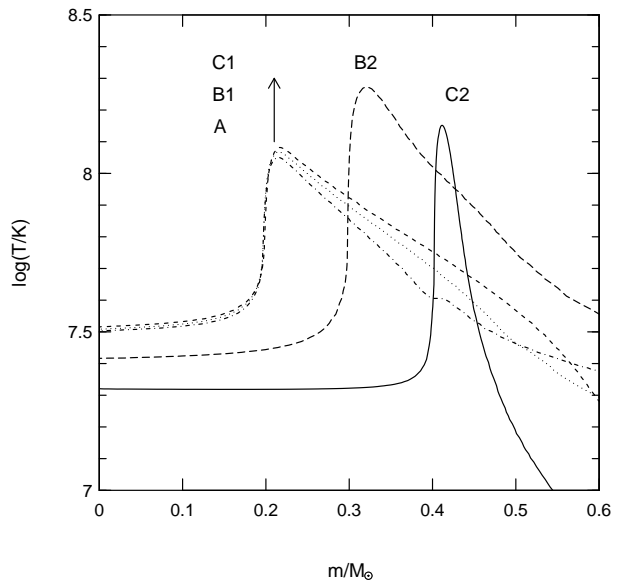}
 \caption{The temperature profile around temperature maximum for five models (labelled,
see text and Table~\ref{table:models}) just after merger. In these models both the RG core and the white dwarf are cold. }
 \label{tprofile}
 \end{figure}

\begin{figure}
 \centering
 \includegraphics [angle=0,scale=0.6]{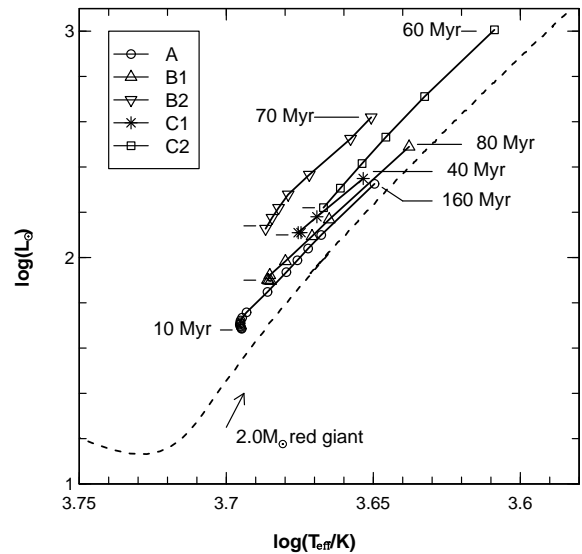}
\caption{The evolutionary tracks of all models. The different symbols with solid
lines indicate the major evolutionary stage for each model (evolution is always
toward higher luminosity). Symbols are separated by intervals of $10^7$yrs. The
first point on each track corresponds to $10^7$yrs after merger (marked by short horizontal
ticks), the last is close to the start of the thermally-pulsing AGB;
both can be identified in Figs.~\ref{02p02cov} (A),
\ref{02p03cov} (B1), \ref{03p02cov} (B2), \ref{02p04cov} (C1), and
\ref{04p02cov} (C2). The dashed line shows the evolutionary track of a normal
2\Msolar\ star.}
 \label{hr}
 \end{figure}

In both "accretion" and "subduction" models, non-degenerate helium is effectively
added to the surface of a degenerate helium core at a high rate. The combination of gravitational
contraction and compression by the outer layers means that a hot shell develops
at the WD/RG-core boundary,  where the maximum
temperature exceeds $10^8K$ (Fig.~\ref{tprofile}). Hydrogen burning starts from the
helium/hydrogen boundary, where the released energy forces the star to expand to large radius
(Fig.~\ref{hr}). The location of helium-ignition depends on the temperature and mass of the RG core
and whether the WD is accreted or subducted. The merged stars are more luminous (or hotter)
than normal red giants because of the larger core masses and, in some cases, an
increased mean
molecular weight in the envelope, due to dredge up of helium and carbon.

In the cases that both the WD and the RG core are cold, a helium-burning flame starts burning from
the merger boundary inward to the center of the core (Fig.~\ref{tprofile}). As the helium-burning
zone moves inwards, there follow a series of helium flashes, each subsequent flash decreasing in
intensity, until the helium-burning shell (flame) reaches the center of the star. During the
He-flash stages, luminosity first increases and then decreases. After the  flame reaches the
center of star, the star goes through a horizontal-branch or red-clump phase (core-helium plus
shell-hydrogen burning). After core-helium exhaustion, a helium-burning shell will again form and
evolution will be similar to the AGB double-shell burning phase.

\begin{figure}
 \centering
 \includegraphics [angle=0,scale=0.6]{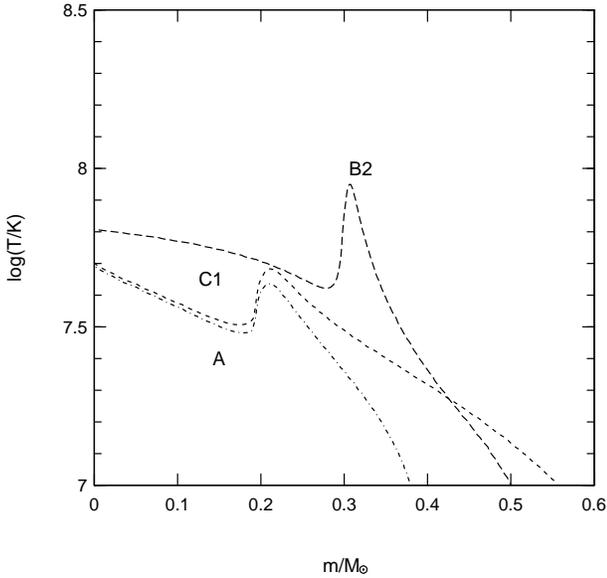}
 \caption{As Fig.~\ref{tprofile}, but for models where the RG core is warm and accretes a cold white dwarf.}
 \label{tpwarm}
 \end{figure}

Whilst our first calculations assumed a cold-degenerate model for the RG core, a more likely
scenario is that the RG core should be {\it warm} at the time of merger (the core is heated by the hydrogen-burning shell). In
practice, the pre-merger temperature of the accreted material is unimportant; fast accretion heats
it to $>10^8$K very quickly. Hence for models where the WD is subducted into the RG core (B1 and C2
below), the approximation is valid. In the remaining cases, we also computed models for accretion
onto a warm RG core (Fig.~\ref{tpwarm}), and report the result for each case below.

The majority of the
test calculations do not show any change in the surface composition of the post-merger; where significant changes occur, we discuss the
results in more detail.

\subsection{Model A: low-mass He WD + low-mass helium core}

\begin{figure}
 \centering
 \includegraphics [angle=0,scale=0.6]{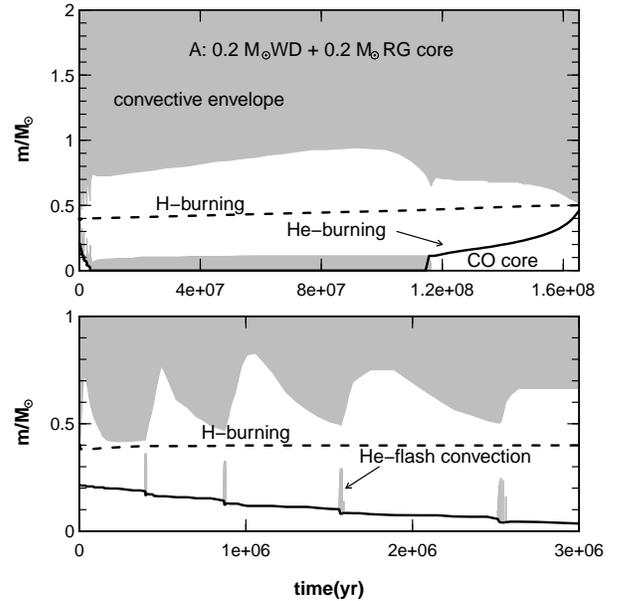}
\caption{The convection structure of Model A, a 0.2\Msolar\ He WD accreted onto
a 0.2\Msolar\ RG core. The grey area shows the convection zones, the dashed line
shows the hydrogen burning front and the solid line shows the helium burning
front. The bottom panel shows an enlargement around the phase of inward helium
burning.}
 \label{02p02cov}
 \end{figure}

Immediately after merger, hydrogen burning drives a fully convective envelope, whilst
 helium burning is ignited at 0.2\Msolar\ (the RG-core boundary) and moves
inwards (Fig.~\ref{02p02cov}). \iso{12}{C} is produced by helium burning through
the $3\alpha$ reaction during
the hot accretion phase. During the inward helium-burning phase, there is no
flash-driven convection zone which can reach the hydrogen-rich envelope. Thus, no burning ash
(\iso{12}{C}) can be dredged up to the surface. In Fig.~\ref{02p02cov}, the grey zone shows
the convection zones, the dashed line shows the hydrogen-burning front and the solid line
shows the helium burning front. Evolution
takes $1.66\times10^8$ yr from merger to the start of the thermally-pulsing  double-shell burning AGB.

In the case of a WD accreted onto a {\it warm} RG core,  helium ignition is nearly central, and more closely
resembles the classical core-helium flash. There is no mixing of processed material to the surface.

These models confirm the \citet{Piersanti2010} conclusion that the merger of a low-mass He WD with a
low-mass helium RG core cannot provide sufficient helium burning to bring \iso{12}{C} to the
envelope.

\subsection{Model B: low-mass He WD + high-mass helium core}
\subsubsection{Model B1 (subduction)}

\begin{figure}
 \centering
 \includegraphics [angle=0,scale=0.6]{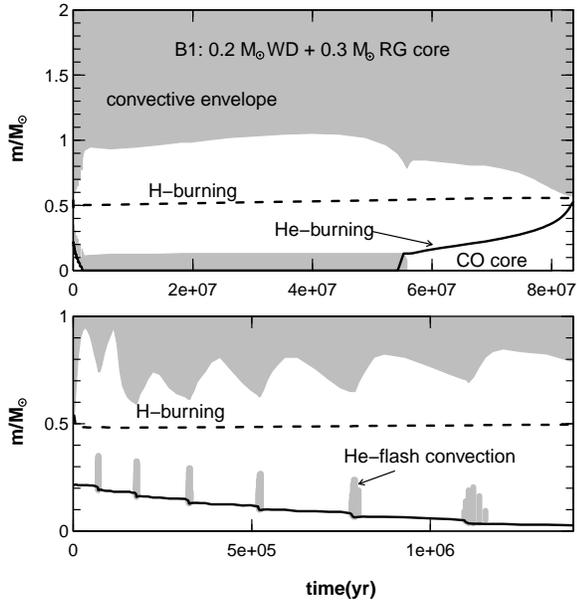}
 \caption{As Fig.~\ref{02p02cov}, but showing the convection structure of Model B1, a 0.2\Msolar\ He WD subducted under a 0.3\Msolar\ RG core. }
 \label{02p03cov}
 \end{figure}

Model B1 (Fig.~\ref{02p03cov}) is similar to model A,  and no flash-driven convection zone reaches the hydrogen envelope. Thus, the evolution is similar except for  helium-core mass (0.5 \Msolar\  compared with  0.4\Msolar\ for case A) and evolution time ($8.4\times10^7$ yr).

\subsubsection{Model B2 (accretion)}

\begin{figure}
 \centering
 \includegraphics [angle=0,scale=0.6]{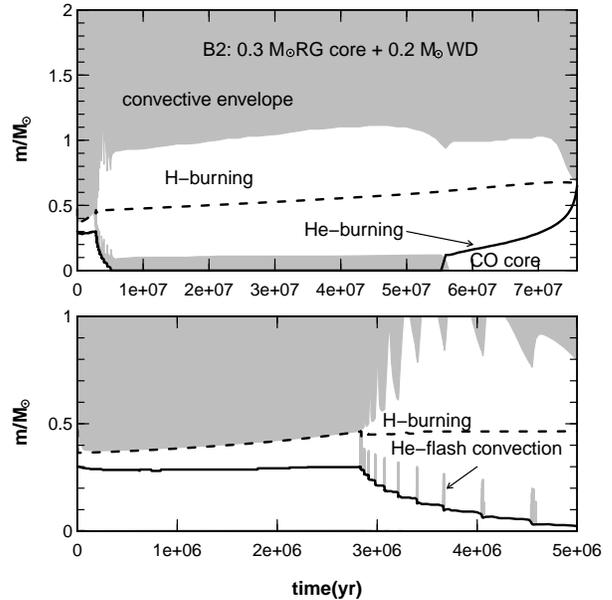}
 \caption{As Fig.~\ref{02p02cov}, but showing the convection structure of Model B2,
a 0.2\Msolar\ He WD accreted onto a 0.3\Msolar\ RG core.}
 \label{03p02cov}
 \end{figure}

Because 0.3\Msolar\ is effectively accreted,  the helium shell in model B2 is much hotter than in
model B1 (where only  0.2\Msolar\ is accreted), especially at the
He/H boundary (Fig.~\ref{tprofile}).  As a consequence,
hydrogen-burning at the boundary is very strong and drives a convection zone which
mixes hydrogen downwards into the helium shell, and also moves the hydrogen-burning
shell downwards by $\approx 0.1$ \Msolar\ immediately after merger (t=0 yr) 
(Fig.~\ref{03p02cov}).  This downwards mixing also dredges up CNO-cycled material
which is mixed throughout  the fully convective envelope,  which thus becomes C-poor and
N-rich.  It takes $7.6\times10^7$ yr from merger to
the thermally-pulsing double-shell burning AGB.

\begin{figure}
 \centering
 \includegraphics [angle=0,scale=0.6]{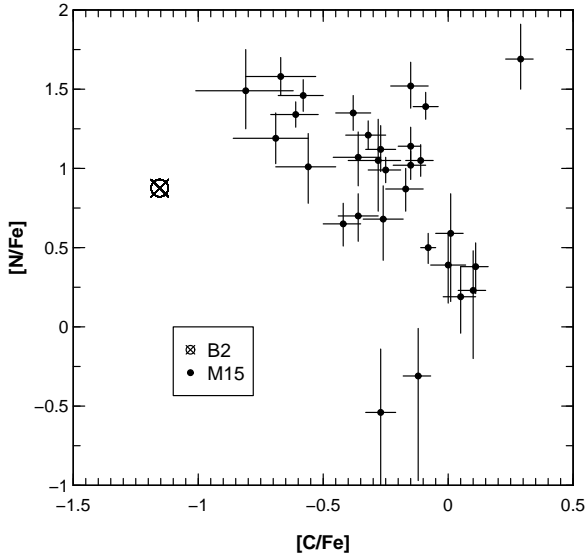}
 \caption{[C/Fe] vs. [N/Fe] ratios for Model B2 compared with the globular stars of M15 (filled circles) \citep{Cohen2005,Bekki2007}. The open circle with cross shows the model abundance of carbon and nitrogen during the major evolutionary stage.}
 \label{03p02ele}
 \end{figure}

There are some globular cluster stars which are C-poor and N-rich which have not
yet been fully explained
\citep{Briley2002,Briley2004,Carretta2005,Cohen2005,Bekki2007}. Our model may
account for some of these stars. Model B2 produces stars with surface
[C/Fe]$=-1.15$, [N/Fe]$=0.88$ and [Na/Fe]$=1.42$, which compares with observed
abundances of [C/Fe], [N/Fe] and [Na/Fe] in the range of $-0.5$ to 0.5, $-1$ to 2
\citep{Cohen2005,Bekki2007} (Fig.~\ref{03p02ele}) and 0.0 to 0.8
\citep{Yong2003,Bekki2007}. Model B2 is also helium rich, with $Y=0.4$; some
fraction of globular cluster stars have $Y>0.3$ \citep{Bekki2007}.

\begin{figure}
 \centering
 \includegraphics [angle=0,scale=0.6]{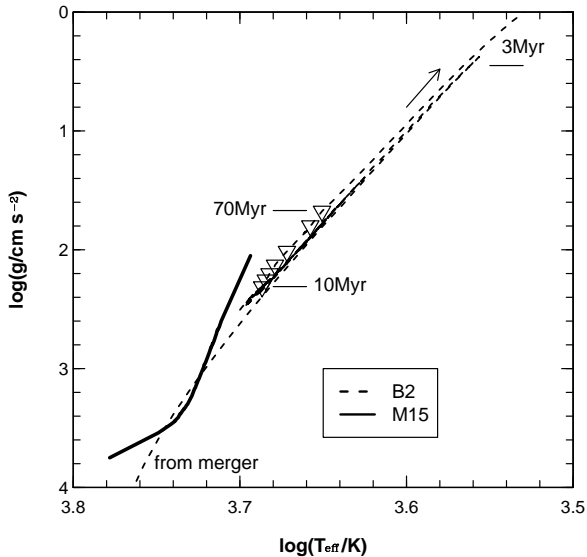}
 \caption{Gravity --  temperature diagram showing the relation adopted for the globular stars of M15 by \citet{Cohen2005,Bekki2007} (thick solid line) and the evolutionary tracks for model B2 after merger (dashed line). The triangles denote intervals of  $10^7$ yrs during the major evolutionary stage (slow core-helium burning); other key times (since merger) are also indicated. }
 \label{03p02gt}
 \end{figure}

The oxygen abundance is in the range observed in globular clusters ({\it e.g.} [O/Fe]=$-0.10$
in our model and $-0.5$ to 0.5 in M15 \citep{Cohen2005,Bekki2007}).
However, our models are more C-poor than most C-poor globular-cluster stars, and
 also, our model produces lower-gravity and lower-temperature stars than observed in M15
(Fig.~\ref{03p02gt}). However, we note that most of these cluster stars have lower metallicity
than our models;  it will be interesting to compute a low metallicity model in the future.

In the case of a WD accreted onto a {\it warm} RG core,  helium ignition is in the same place
(the WD / RG-core boundary) but, because the core is non-degenerate and
the maximum temperature at the WD / RG-core boundary is {\it lower} than that in the cold-core model
(Fig.\,\ref{tpwarm}),
the shell flash is not strong enough to drive convection
through the hydrogen shell, and there is {\it no} mixing of CNO-processed material to the surface.

\subsection{Model C: high-mass He WD + low-mass helium core}

\subsubsection{Model C1 (accretion)}

\begin{figure}
 \centering
 \includegraphics [angle=0,scale=0.6]{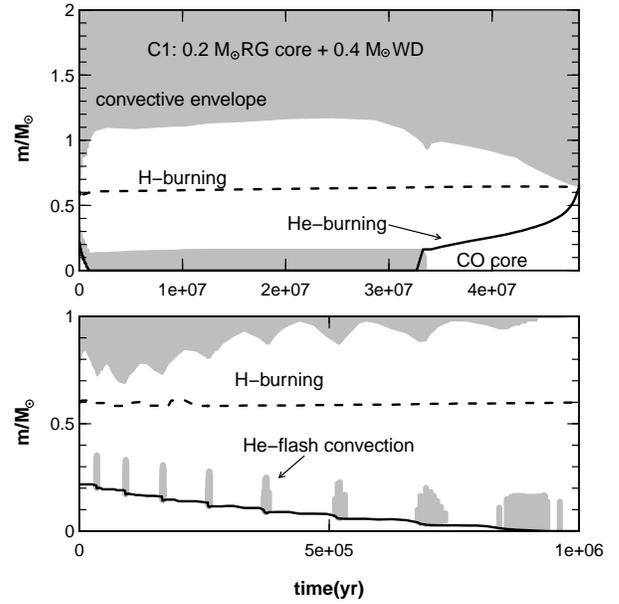}
\caption{As Fig.~\ref{02p02cov}, but showing the convection structure of Model
C1, a 0.4\Msolar\ He WD accreted by a 0.2\Msolar\ RG core.}
 \label{02p04cov}
 \end{figure}

Model C1 (Fig.~\ref{02p04cov}) is similar to Models A and B1 and, again, no flash-driven convection reaches the hydrogen envelope.
Thus, evolution is similar except for a different core mass and evolution time ($4.8\times10^7$ yr).
The result for a WD accreted onto a {\it warm} RG core is  also similar to that for case A, with no surface mixing.

\subsubsection{Model C2 (subduction)}

 \begin{figure}
  \centering
  \includegraphics [angle=0,scale=0.6]{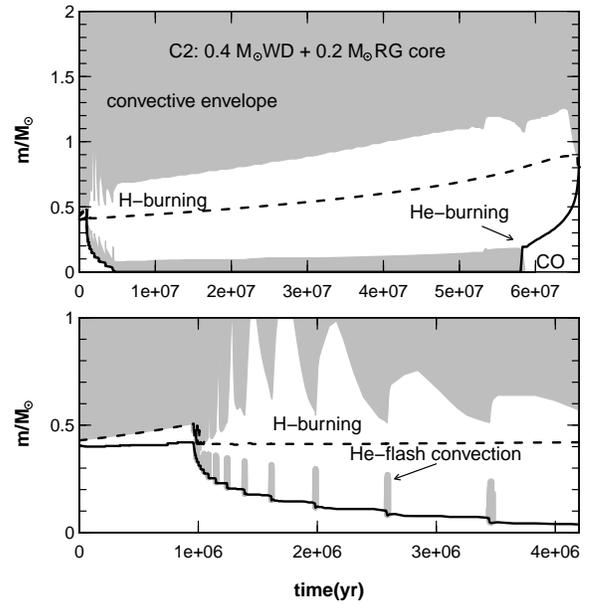}
  \caption{As Fig.~\ref{02p02cov}, but showing the convection structure of Model C2, a 0.4\Msolar\ He WD subducted beneath a 0.2\Msolar\ RG core. }
  \label{04p02cov}
 \end{figure}

In model C2 (Fig.~\ref{04p02cov}), the helium shell is very hot,
and enhanced hydrogen-burning mixes $\approx 0.2$\Msolar\  hydrogen into the helium shell.
At the beginning of evolution, hydrogen burning forms a fully convective envelope, and
forces the star to expand.
 Immediately following merger (and for approximately 150 years thereafter),
the hydrogen-burning shell converts available  \iso{12}{C} to  \iso{13}{C}  and \iso{14}{N},
thus depleting \iso{12}{C} in the convective envelope and decreasing the  \iso{12}{C} /\iso{13}{C}
ratio by a factor of $\approx 0.2$ (the   \iso{12}{C} /\iso{13}{C} ratio in the envelope of the RGB progenitor is 26).
Since the initial mixing is short-lived (50 years or four timesteps in our model), this factor is likely
to be subject to several sources of uncertainty including the convection physics, numerics and
the model parameters ({\it e.g.} masses and accretion rates).

After this, the hydrogen-burning shell detaches from the convection zone.
Meanwhile, the core is relatively cold (20 MK), so the helium-burning shell is weak
and relatively inactive ($L_{\rm He}/L_{\rm H}\approx 10^{-6}$),
whilst the hydrogen shell burns outwards through  $\approx 0.1$\Msolar.
After $10^6$yr, the He-shell temperature rises sufficiently for a helium-shell
flash to occur, driving a convection zone which mixes through to the hydrogen-rich envelope.
Helium-burning continues  in a series of flashes from the merger
boundary to the center of the core.

\begin{figure}
\centering
\includegraphics [angle=0,scale=0.6]{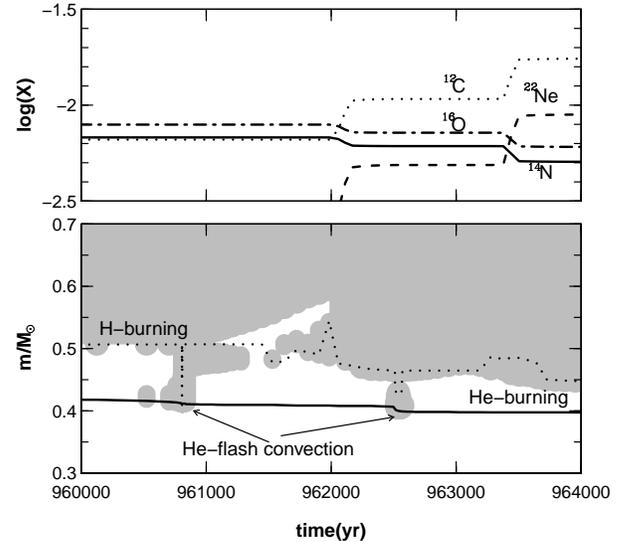}
\caption{Bottom: as Fig.~\ref{04p02cov}, but expanded around the the first two helium-shell
flashes of Model C2. Top: the evolution of the surface abundance of \iso{12}{C}, \iso{14}{N}, \iso{16}{O}
and \iso{22}{Ne} over the same interval. }
\label{zoomc2cov}
\end{figure}

During the first two helium-shell flashes, flash-driven
convection connects the helium-burning shell to the convective
envelope and dredges freshly-produced   \iso{12}{C}  ash  from the
helium-burning shell to the surface
(the duration of these mixing events is short ($<200$ yr) so that insufficient
$\iso{12}{C}(p,\gamma)\iso{13}{N}(,\beta^+\mu)\iso{13}{C}(p,\gamma)\iso{14}{N}$
reactions occur as the carbon transits the H-shell to affect the surface
\iso{13}{C} or \iso{14}{N} abundances directly). However, the  events
are long enough compared with the convective turnover time,
assuming that this is comparable with a normal red-giant envelope,
that the envelope will be fully mixed. 
The evolution of the structure is shown in Fig.~\ref{zoomc2cov} (lower panel),
where the position of the H-burning shell is indicated by the maximum in
nuclear energy production, which is very susceptible to local mixing events.
Fig.~\ref{zoomc2cov} also shows the surface-abundance evolution of
several species over the same interval (upper panel), demonstrating the dredge-up of
\iso{12}{C} from 3$\alpha$ burning, enhanced \iso{14}{N} from CNO-cycle helium burning
([N/Fe]=0.69), \iso{22}{Ne}
from $\alpha$ captures on \iso{14}{N}, and the depletion of \iso{16}{O}, also associated with
the CNO-cycle.  The detailed evolution of the structures is likely to be sensitive to
resolution and time step, as is the stability of the calculation; numerical problems
with forced changes to the time step prevented further exploration.

\begin{figure}
  \centering
  \includegraphics [angle=0,scale=0.6]{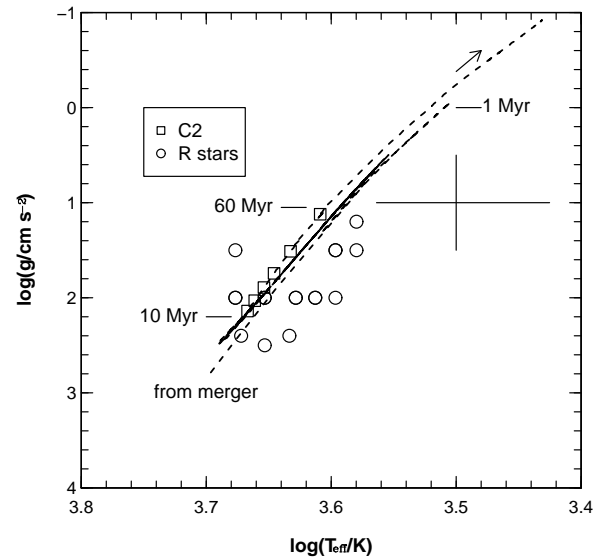}
  \caption{Gravity --  temperature diagram showing  the early-type R stars (circles) \citet{Zamora2009}
and   the evolutionary tracks for model C2 (dashed line). Squares denote intervals of $10^7$ yrs during the major evolutionary stage. The cross shows the mean errors for the  R-star measurements.}
  \label{04p02gt}
 \end{figure}

The model takes $6.6\times10^7$ yr to evolve from merger to
the thermally-pulsing double-shell burning AGB;  most of the time is spent as a red-clump
star on the giant branch (squares in Fig.~\ref{04p02gt}.)

 \begin{figure}
 \centering
 \includegraphics [angle=0,scale=0.6]{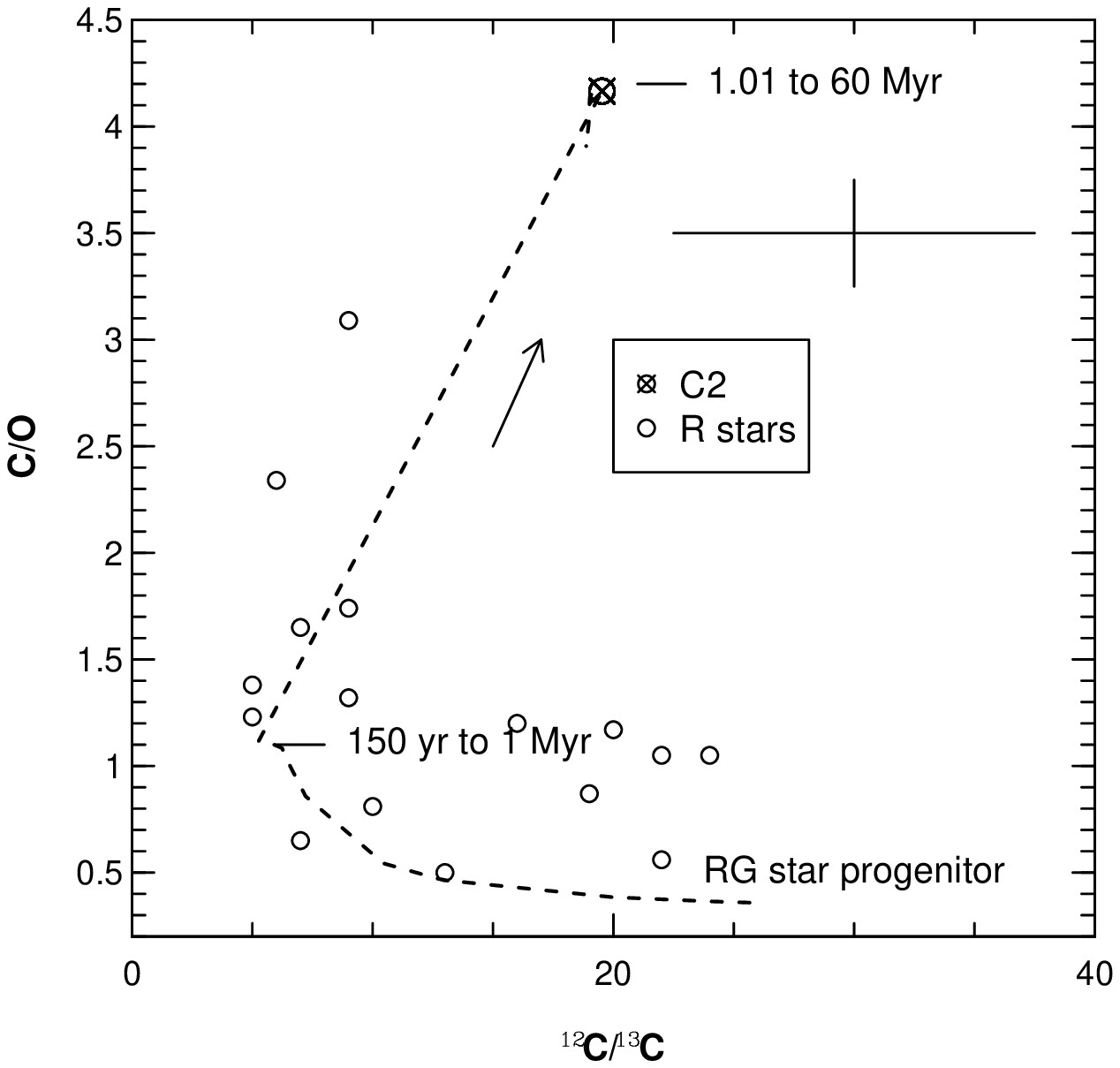}
\caption{The post-merger evolution of the C/O and \iso{12}{C}/\iso{13}{C} ratios for Model C2 compared with
the early-R stars of \citet{Zamora2009}. The star HIP 58 786 has
\iso{12}{C}/\iso{13}{C}=70 and is not shown on this diagram. The dashed line
shows the model evolution of C/O and \iso{12}{C}/\iso{13}{C} ratios after
merger. The circle with cross shows the ratios of the C/O and \iso{12}{C}/\iso{13}{C} during
most of this evolution. The cross shows the mean errors for the  R-star measurements.} \label{04p02ele}
 \end{figure}

Fig.~\ref{04p02gt} compares model C2 with 15 early-R stars from \citet{Zamora2009} on a
$g-T_{\rm eff}$ diagram.  The  observed stars are satisfactorily close to the evolutionary tracks.
However,  since most of our models (and models of other giants) are almost coincident on such
a diagram, this is more of a consistency check than a strong constraint.
 Fig.~\ref{04p02ele} compares the C/O ratios and  \iso{12}{C}/\iso{13}{C} ratios for Model C2
with the early-R star measurements \citep{Zamora2009}. During the first million years after
merger,  the model evolves through the region occupied by early-R stars. The precise trajectory
is subject to the uncertainty in the initial depletion of \iso{12}{C} discussed above, but the start and
end points should be similar;
it is conjectured that the allowable trajectory envelope should bracket the loci of observed early-R stars.
The models generate a similar abundance of C and O to that observed.
Furthermore, we find that \iso{22}{Ne} and \iso{23}{Na} are enriched to give
[\iso{22}{Ne}/Fe]=1.98 (\iso{20}{Ne}/\iso{22}{Ne}= 0.17) and [Na/Fe]=1.46.
 It is not known whether early-R stars are enriched in these species.
Although the \iso{22}{Ne}($\alpha$,n)\iso{25}{Mg} is a putative neutron source
for further (s-process) nucleosynthesis \citep{Cameron1960},  conditions
in our models ($T_{\rm max} \approx 1.4 \times 10^8 {\rm K}$  with
$\rho \approx 3.8 \times 10^3 {\rm g/cm}^3$ in the helium shell) do not satisfy
the neccessary criteria ($T > 2.2-3.5 \times 10^8 {\rm K}$
and $\rho > 1-3\times 10^3 {\rm g/cm}^3$)
\citep{Raiteri1991a,Meyer1994}. Moreover, s-process nucleosynthesis
is not included in the version of MESA used in these calculations.  

For Model C2 with a final mass of 1.5\Msolar\, the evolutionary tracks are similar to those with a final mass of 2.0\Msolar,  except that the final C/O ratio increases to 6.1 (instead of 4.2).
This is because both models have the same core mass, and dredge up the same mass of carbon, which is
consequently more concentrated in the lower-mass envelope.

\begin{figure}
 \centering
 \includegraphics [angle=0,scale=0.6]{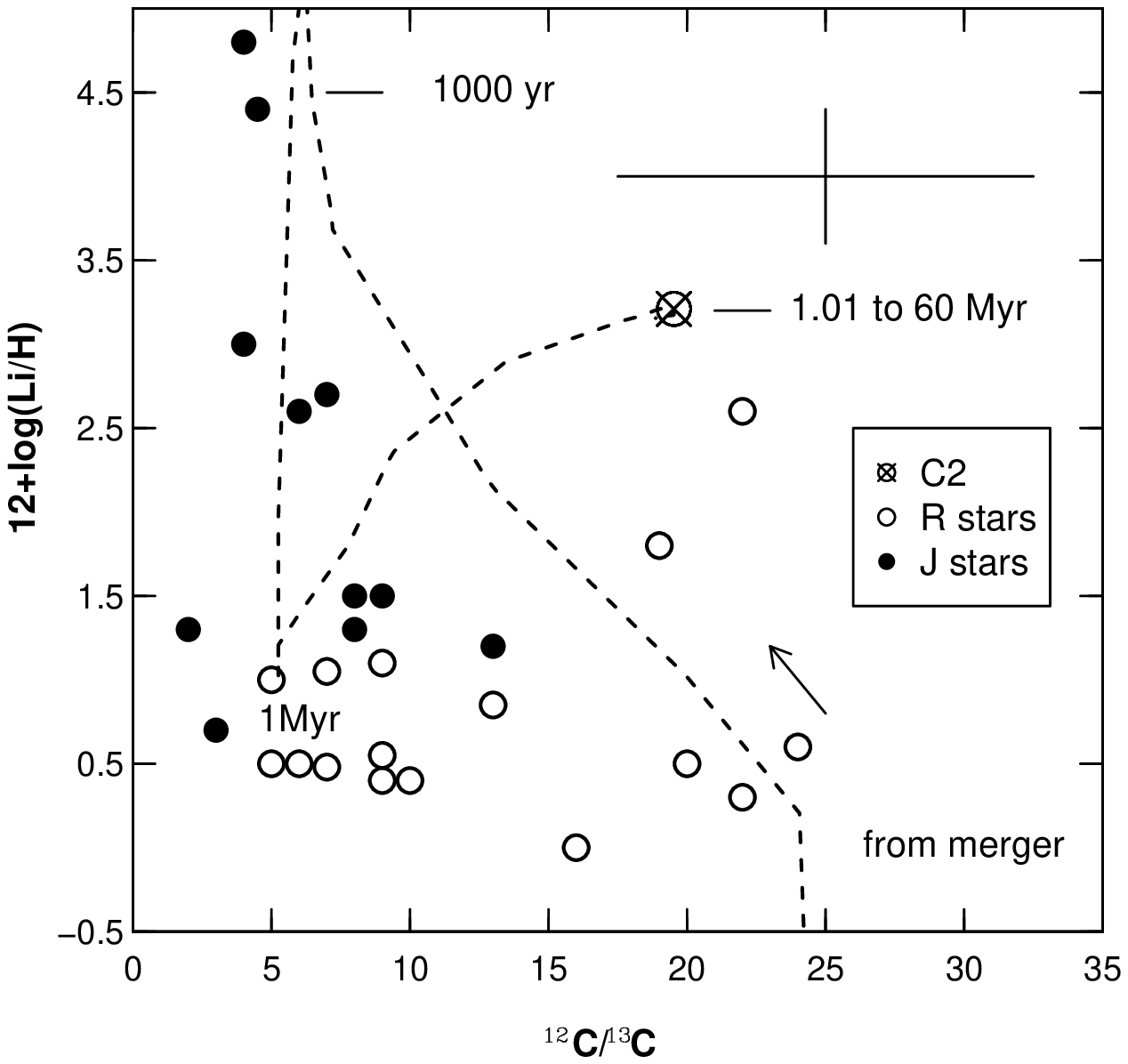}
 \caption{Lithium versus \iso{12}{C}/\iso{13}{C}. Circles show the early-R stars from \citet{Zamora2009}. Dots show the J stars from \citet{Abia2000}. The dashed line shows the Li/H and \iso{12}{C}/\iso{13}{C} ratios during the evolution of model C2 immediately after merger. The circle with cross shows the Li/H and \iso{12}{C}/\iso{13}{C} ratios during the main core-helium burning phase. The cross shows the mean errors for the  R-star measurements.}
 \label{04p02li}
 \end{figure}

\subsection{J stars and Lithium}

Early-R and J stars have a notably large overabundance of lithium,
which has so far been difficult to explain. To explain the overabundance of lithium in R\,CrB stars, \citet{Longland12} showed that lithium can be produced by the merging of a helium white dwarf with a carbon-oxygen white dwarf if their chemical
composition is rich in \iso{3}{He} from the previous evolution. \citet{Zhang12b} confirmed this in their He+He WD merger simulation. This
model requires enough \iso{3}{He} to be left in the white dwarf after the end of
main-sequence evolution and a hot enough zone to form during the merger.
In our simulation, we obtained a mass fraction of $5.6\times10^{-4}$ \iso{3}{He} in the RG envelope  before merger.
We introduced this mass fraction into the envelope of the post-merger models. The hot He-burning shell drives convection which mixes
$\iso{3}{He}$ down into the He-burning shell where it burns with $\iso{4}{He}$ to produce $\iso{7}{Li}$ [$\iso{3}{He}(\iso{4}{He},\gamma)\iso{7}{Be}(e^-,\nu)\iso{7}{Li}$].

\begin{figure}
 \centering
 \includegraphics [angle=0,scale=0.6]{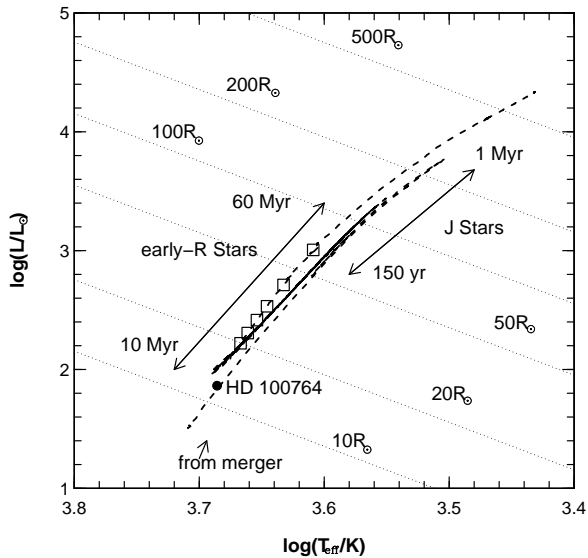}
 \caption{The evolutionary track of Model C2. Squares represent intervals of $10^7$yrs.
The double-ended arrows show the  luminosity ranges for model C2 over the time
intervals indicated on the figure. These correspond approximately with the
observed luminosity ranges of early-R and J stars.
The dot shows HD 100764.}
 \label{rj}
 \end{figure}

For the new born \iso{7}{Li} to survive, the \iso{7}{Be} must be transferred to a cooler region
rapidly before the decay occurs. Thus, in a manner similar to that described by \citet{Cameron71},
the \iso{7}{Li} may or may not appear on the surface depending on whether sufficient convection
occurs (see Fig.~\ref{04p02cov}). In our models, Models B2 and C2 both produce lithium. However, in
Model B2 the new-born lithium is almost totally destroyed immediately by the reaction
$\iso{7}{Li}(p,\iso{4}{He})\iso{4}{He}$. In Model C2, the star retains its high lithium abundance
because the mixing is very effective.
While the star goes through its high-luminosity phase, its
surface reaches a peak lithium abundance while the \iso{12}{C}/\iso{13}{C} ratio is still low
(Fig.~\ref{04p02li}). Thus, J stars may represent just a special stage during the formation of
early-R stars, as already hinted at by \citet{Abia2000}.

This stage, where \iso{12}{C}/\iso{13}{C} is very low, takes around 1 Myr, compared
with 60 Myrs for the lifetime of Model C2. If C2 represents early-R stars, then J stars should
constitute $\approx1/60$ of the total (Fig.~\ref{rj}).

Although the statistical evidence is not very strong, there are believed to
be around 10 times more early-R stars than N stars  \citep{Blanco65}.
If the J stars constitute $\approx1/60$ of early-R stars,
then the J:N ratio would be $\approx 15\%$ which is the observed value \citep{Abia2000,Blanco65}.

This argument assumes that J-stars, like early-R stars, are all single. This may not be true.

Approximately 10-15\% of J stars are believed to have a silicate dust shell \citep{Lloyd91}. 
Some consider
such shells to be circumbinary disks. Although there is evidence for disk asymmetry,
no J-star companion, or J-star orbit has  so far been established \citep{Deroo07}. 
 If we consider
the J-star phase of Model C2 to represent the immediate post-merger phase, it is possible that the
observe silicate dust shell may form from the ejected part of the envelope.
 Since energy considerations demand that some fraction of the
envelope should be ejected during merger,  an asymmetric shell (or disk) would be an obvious
geometry for the ejecta.
If a silicate dust shell forms during merger and has a
lifetime $\approx 10^5$ years, then we might expect  a fraction of
10-15\% to show such shells, since the J stars are younger than the majority of early-R stars and
have a lifetime of $\approx 1$Myr.
We still need to explain why HD100764, an early-R star  (Fig.~\ref{rj}),  also
shows a dust shell \citep{Parthasarathy91,Sloan2007}.
However it has quite a low ratio of \iso{12}{C}/\iso{13}{C}=4, and shows
emissions from polycyclic aromatic hydrocarbons (PAHs) \citep{Sloan2007}. We therefore suggest that
HD100764 might be a very recent merger currently evolving {\it towards} the J-star phase.

\section{Conclusion}

Mergers of helium white dwarfs with red giants have been considered as a possible model for the
origin of early-R stars. In this paper we have investigated three possible channels for He WD +
RG mergers, and find that only a high-mass He WD subducting into a low-core-mass RG star (Model C2)
can produce early-R stars. However, nucleosynthesis of elements C, N, O, and Li for this model
match the observations well. Moreover, J-stars may represent a short stage in this evolution. The
fact that only Model C2 can produce early-R stars may explain why \citet{Izzard2007} produce ten
times as many early-R stars as required. Our study shows that two of their proposed merger channels
do not produce the correct surface composition. A more detailed population-synthesis study should be
carried out to distinguish the boundaries of each of the evolution channels.

Our study has also shown that a low-mass He WD accreted onto a high-mass RG core may account for
some C-poor and N-rich globular cluster stars. Lithium enrichment in red giants has previously been
explained by one or more of rotational mixing, mass loss, planet accretion and deep circulation
\citep{Charbonnel00}. In this paper we have shown that a He WD + RG merger may be an additional and
important channel for producing lithium-rich red giants.

\section*{Acknowledgments}
The Armagh Observatory is supported by a grant from the Northern Ireland Dept. of Culture Arts and Leisure.

\bibliographystyle{mn}	
\bibliography{mybib}		

\label{lastpage}

\end{document}